\documentclass[12pt,a4paper]{article}
\usepackage{graphicx,amssymb,amsmath}
\usepackage{setspace} 
\usepackage[linkcolor={blue},citecolor={blue},colorlinks=true,linktocpage,urlcolor={blue}]
{hyperref}
\usepackage{cite}
\usepackage[margin=2cm]{geometry}
\usepackage[dvipsnames]{xcolor}
\definecolor{darkblue}{rgb}{0,0,0.54}
\usepackage{multirow} 
\usepackage{appendix}
\usepackage{pdfpages}
\usepackage{tabu} 
\usepackage{adjustbox} 
\usepackage{braket}
\usepackage[skip=2pt,font=scriptsize]{caption}
\usepackage{color,soul}
\usepackage[colorinlistoftodos]{todonotes}
\usepackage[normalem]{ulem}
\usepackage{authblk}

\usepackage{bm} 
\usepackage[symbol]{footmisc}

    \makeatletter
\def\@fnsymbol#1{\ensuremath{\ifcase#1\or \dagger\or \ddagger\or
   \mathsection\or \mathparagraph\or \|\or **\or \dagger\dagger
   \or \ddagger\ddagger \else\@ctrerr\fi}}
    \makeatother
\graphicspath{{Figures/}} 

\begin{document}


\title{Sub–cycle
field–driven dynamical Berry phase in solids}

\global\long\def\ket#1{\left|#1\right\rangle }%
\global\long\def\bra#1{\left\langle #1\right|}%
\global\long\def\braket#1#2{\left\langle #1\right|\left.#2\right\rangle }%
\global\long\def\at#1{\left.#1\right|}%

 \author[1]{Lior Faeyrman}
 \author[2]{Jianing Zhang}
 \author[3,4]{Misha Ivanov}
 \author[2,5,6]{Liang-You Peng}
 \author[1,$^{\dagger}$]{Nirit Dudovich}
 \author[7,$^{*}$]{Riccardo Piccoli}



	\affil[1]{Department of Complex Systems, Weizmann Institute of Science, 76100 Rehovot, Israel}

    \affil[2]{State Key Laboratory for Mesoscopic Physics and Frontiers Science Center for Nano-optoelectronics, School of Physics,
Peking University, Beijing 100871, China}

\affil[3]{Max-Born-Institut, Max-Born Strasse 2A, D-12489 Berlin, Germany}
	
    \affil[4]{Technische Universit\"at Berlin, Ernst-Ruska-Geb\"aude, Hardenbergstr. 36A, D-10623 Berlin, Germany}

    \affil[5]{Beijing Academy of Quantum Information Sciences, Beijing 100193, China}

\affil[6]{Collaborative Innovation Center of Extreme Optics, Shanxi University, Taiyuan 030006, China}

\affil[7]{Department of Molecular Sciences and Nanosystems, Ca' Foscari University of Venice, 30172 Venice, Italy}



\def\thefootnote{$\dagger$}\footnotetext{Corresponding author: nirit.dudovich@weizmann.ac.il}\def\thefootnote{\arabic{footnote}}
\def\thefootnote{*}\footnotetext{Corresponding author: riccardo.piccoli@unive.it}\def\thefootnote{\arabic{footnote}}

\maketitle

\begin{abstract}
In quantum mechanics, a wavepacket acquires a geometric phase -- known as the Berry phase -- as it evolves along a closed trajectory in parameter space \cite{berry_quantal_1984}. In condensed matter systems, the Berry phase underlies a broad range of phenomena, including the anomalous Hall effect, orbital magnetism, and electric polarization \cite{xiao_berry_2010}. However, in centrosymmetric materials possessing time-reversal (TR) symmetry, its manifestation is suppressed and effectively vanishes. When a system is driven by a strong terahertz (THz) field, it can be coherently driven far from equilibrium, transiently reshaping its symmetry on sub-picosecond timescales \cite{kampfrath2013resonant,SALEN20191}. This capability opens new avenues for quantum control with potential applications in information processing and sensing. Here, we experimentally demonstrate that a strong THz field can transiently break inversion symmetry in MgO, inducing a dynamical complex Berry phase, thereby manipulating the topological properties of the material. Applying high-harmonic generation (HHG) spectroscopy, we directly resolve the Berry phase, accessing both its real and imaginary components. The first is associated with coherent intraband dynamics \cite{uzan-narovlansky_observation_2024} while the second with quantum tunneling through a potential barrier \cite{faeyrman2024revealing}. This observation enables the reconstruction of the time-dependent evolution of the Berry phase within the cycle of the THz field. The coherent manipulation of solids with strong fields, combined with attosecond-resolved HHG spectroscopy, represents a fundamental step toward unveiling and controlling geometric quantum phenomena in condensed matter systems.

\end{abstract}


\newpage
\section{Main}

Whenever a fast-oscillating system evolves over a slowly oscillating parameter space, a geometric phase may arise, also known as the Berry phase. This concept is deeply rooted in a wide range of fields in physics, including classical physics    \cite{von_bergmann_foucault_2007}, optics \cite{pancharatnam_generalized_1956,simon1988evolving}, quantum mechanics \cite{berry_quantal_1984}, condensed matter physics \cite{xiao_berry_2010}, fluid mechanics \cite{berry_wavefront_1980}, and particle physics \cite{wilson1974confinement}. 
Previous studies have shown that the Berry phase can consist of both real and imaginary components, corresponding to geometric phase and amplitude modulation, respectively \cite{berry1990geometric,zwanziger1991measuring}. The Berry phase is not merely a theoretical construct, but it has had profound implications in condensed matter physics. It underlies the modern theory of polarization in crystals \cite{king-smith_theory_1993}, plays a central role in the topological classification of materials \cite{hasan2010colloquium}, and is fundamental to the understanding of the anomalous Hall effect \cite{thouless1982quantized}. 
Although the Berry phase plays a central role in many physical phenomena, it does not naturally arise in all systems. Its emergence requires specific symmetry-breaking conditions. In systems that preserve both time-reversal (TR) and inversion symmetry, the Berry curvature vanishes. In such systems, the Berry phase, which corresponds to the surface integral of the curvature over the region enclosed by the closed path, is absent \cite{xiao_berry_2010}. Light-matter interactions with these systems raise fundamental questions: Can we transiently break one of the symmetries that protect the system and induce a dynamical Berry phase? How can we resolve the temporal evolution of this phase?


In recent years, the active control of topological properties of matter has emerged as one of the most important frontiers in quantum materials research, spanning field-driven valleytronics \cite{tyulnev_valleytronics_2024, mitra2024light, jimenez2020lightwave}, magnetic control in graphene \cite{doi:10.1126/science.aal0212}, electrical switching in topological insulators \cite{xu2018electrically}, and light-induced modulation of anomalous Hall conductivity in Weyl semimetals \cite{yoshikawa2022non}. While these studies have primarily focused on manipulating specific topological properties of quantum materials, the field-induced complex Berry phase and its temporal evolution during transient symmetry breaking have remained experimentally inaccessible.

 

In this study, we present and experimentally demonstrate the field-induced generation of a complex Berry phase in MgO as a prototypical wide-bandgap (7.8 eV) centrosymmetric insulator. By applying an intense terahertz (THz) field, we dynamically break the crystal’s inversion symmetry (see Fig. 1a), inducing a field-driven Berry curvature. Driving the electronic wavefunction with a femtosecond infrared (IR) pulse induces the accumulation of a Berry phase within a fraction of the IR cycle. We resolve, with attosecond precision, both the real and imaginary components of the interband Berry phase using two-color interferometry \cite{uzan-narovlansky_observation_2024}. 
Resolving the temporal evolution of this phase reveals its modulation within the cycle of the driving THz field.

We demonstrate field-induced Berry phase by employing a high-field THz source, previously applied to drive lattice vibrations, control spin dynamics, access hidden phases, and induce transient symmetry breaking \cite{kampfrath2013resonant,SALEN20191}. Such THz fields offer unique advantages: their long wavelengths ($\sim$mm) and low photon energies ($\sim$meV) allow for symmetry control, while minimizing heating, material damage, and ionization. The ability to break symmetry by applying THz fields was first demonstrated through second-harmonic generation (SHG) in various systems \cite{nahata1998detection,cook1999terahertz,grishunin2017thz}. Recently, Vampa et al. \cite{Vampa2018} and Li et al. \cite{Li2023} demonstrated that high harmonic generation (HHG) spectroscopy can serve as an accurate probe of such symmetry breaking, giving rise to even-order harmonics, while Liu et al. \cite{liu2025terahertz} theoretically studied THz-modulated HHG in a two-dimensional material. Building on this foundation, we take a significant step forward by inducing a topological feature -- the dynamical complex Berry phase -- and tracking its sub-cycle evolution.

 The interaction between a strong THz field and a wide-bandgap material serves as a powerful ``knob'' for dynamically controlling the material’s topological properties. Resolving this field-induced topology is based on the integration between multiple temporal scales that differ by several orders of magnitude. HHG spectroscopy provides access to ultrafast electronic and topological dynamics evolving on attosecond timescales. The driving IR field, spanning tens of femtoseconds, serves as a narrow temporal gate that probes the material’s response within the significantly slower THz cycle, which unfolds over picoseconds. This multi-scale framework enables us to temporally resolve the evolution of field-induced Berry phase with sub-cycle precision.

Consider a wide-bandgap material that is both centrosymmetric and TR symmetric. When driven by a strong, linearly polarized IR field with frequency $\omega_{\mathrm{IR}} \ll \varepsilon_g$ (where $\varepsilon_g$ is the bandgap, in atomic units), the generated high harmonics contain only odd orders, due to the symmetry of the interaction. We apply a THz field parallel to the IR drive, having an amplitude sufficiently strong to break inversion symmetry, while remaining perturbative with respect to the bandgap, where $\omega_{\mathrm{THz}} \ll \omega_{\mathrm{IR}} \ll \varepsilon_g$. In this case, breaking the unit cell symmetry generates an intrinsic dipole moment and induces a Berry connection that, to leading order, is proportional to the applied THz field (see SI for details). Due to the large separation of timescales between the slow THz field and the fast electronic response, the injected electron experiences a transient crystal potential that lacks inversion symmetry. As a result, the high-harmonic spectrum now contains both even and odd orders. Crucially, this THz-modulated crystal acquires topological features -- such as a nontrivial Berry phase -- that are absent in the equilibrium material. 


To capture the transient emergence of the Berry phase, we use the exceptional temporal resolution offered by solid-state HHG spectroscopy \cite{ghimire2011observation,vampa2017merge}, revealing underlying dynamics on their natural timescales. A key advantage of this scheme lies in the attosecond scale of the nonlinear process. It maps the coherence of the quantum wavefunction before scattering or dephasing mechanisms set in, thereby preserving the quantum nature of the system that would otherwise be obscured at longer timescales. In wide band-gap materials, such as MgO, where the band gap exceeds the photon energy of the driving field, the HHG process is governed by the coexistence of interband and intraband currents. Under a strong IR field, the electronic wavefunction can tunnel across the band gap. Driven by the field, it propagates through the conduction bands, emitting intraband radiation, and may eventually recombine with its parent hole, transferring its accumulated energy to the emitted harmonics in a process known as interband HHG \cite{vampa2014theoretical,vampa2015linking,chacon2020circular}. Previous studies, along with our simulation results, demonstrate that under our experimental conditions the harmonic emission is dominated by interband currents, thereby supporting an electron-trajectory picture of the HHG process (see SI for details).

In MgO, the dipole coupling is governed primarily by a single valence band, allowing an effective two-band approximation to capture the essential physics of the system\cite{uzan_attosecond_2020,uzan-narovlansky_observation_2022}. The validity of this simplification is confirmed by comprehensive multi-band numerical simulations (see SI for details). Within this theoretical framework, and in the absence of the THz perturbation, the emitted radiation can be described using the well-established Saddle Point Approximation (SPA) to the integral form of the semicoductor Bloch equations (SBE) \cite{lewenstein_theory_1994,chacon2020circular,vampa2017merge}. In this case, the centrosymmetric semiclassical action is given by $S_{0}\left[\boldsymbol{k_0},t_0,t\right]=\int_{t_{0}}^{t}\varepsilon_{g}\left[\boldsymbol{k}\left(t'\right)\right]dt'+\left[\phi_{d}\left(\boldsymbol{k}\left(t_{0}\right)\right)-\phi_{d}\left(\boldsymbol{k}\left(t\right)\right)\right]$,
where $\left[\boldsymbol{k_0},t_0,t\right]$ are the saddle point solutions for ionization momentum, ionization time, and recombination time, respectively. $\varepsilon_g=\varepsilon_{c}-\varepsilon_{v}$ is the conduction-valence band-gap, $\boldsymbol{k}\left(\tau\right) =\boldsymbol{k_0}-\boldsymbol{A}\left(t_{0}\right)+\boldsymbol{A}\left(\tau\right)$ the crystal quasi-momentum with $\boldsymbol{A}(t)$ the IR field vector potential, and $\phi_d=i\arg\!\bigl\{\bra{u_{v\boldsymbol{k}}}\nabla_{\boldsymbol{k}}\ket{u_{c\boldsymbol{k}}}\bigr\}$ the conduction-valence transition dipole moment phase with $\ket{u_{n\boldsymbol{k}}}$ the periodic part of the crystal's Bloch function of energy band $n$ and crystal momentum $\boldsymbol{k}$.

In the absence of the THz field, the integral of the Berry connection $\boldsymbol{\mathcal{A}}_{n}=i\langle u_{n\boldsymbol{k}}|\nabla_{\boldsymbol{k}}|u_{n\boldsymbol{k}}\rangle$ over the Brillouin zone (BZ) in a centrosymmetric material, $\int_{BZ}\boldsymbol{\mathcal{A}_n}\cdot d\boldsymbol{k}$, vanishes (modulo a lattice vector) due to symmetry \cite{king-smith_theory_1993}. Physically, this corresponds to the Wannier center of the electronic wavefunction being located at the inversion center of the unit cell \cite{silva2019high,kitamura2020nonreciprocal}. Adding a THz field introduces a perturbation to the semiclassical action, modifying it as $S_{0}\rightarrow S_{0}+\sigma_{THz}$. This modification
alters the eigenstates of the system $|u_{n\boldsymbol{k}}\rangle\rightarrow|u'_{n\boldsymbol{k}}\rangle$ and, to first order in the THz field, a non-vanishing Berry connection is induced, namely $\boldsymbol{\mathcal{A}}'_{n}=i\langle u'_{n\boldsymbol{k}}|\nabla_{\boldsymbol{k}}|u'_{n\boldsymbol{k}}\rangle\propto\boldsymbol{F}_{THz}$ with $\boldsymbol{F}_{THz}$ being the THz electric field vector. Importantly, no gauge can be chosen to nullify it across the entire BZ (see SI for the full derivation). This effect can be interpreted as a displacement of the Wannier center to a non-trivial position within the unit cell, breaking the internal wavefunction symmetry. As a result, the effective band-gap is modified as: $\varepsilon_g\rightarrow\varepsilon'_{g}=\varepsilon_{g}+\boldsymbol{F}_{IR}\cdot\left(\boldsymbol{\mathcal{A}}^\prime_g+\nabla_{\boldsymbol{k}}\phi_d\right)$ with $\boldsymbol{F}_{IR}$ the fundamental electric field vector, and $\boldsymbol{\mathcal{A}}'_{g}=\boldsymbol{\mathcal{A}}'_{c}-\boldsymbol{\mathcal{A}}'_{v}$ the induced conduction-valence Berry connection\cite{chacon2020circular}. Finally, the modified action now reads:

\begin{equation}
    S[\boldsymbol{k}_0,t_0,t]=\int_{t_{0}}^{t}\left[\varepsilon_{g}\left[\boldsymbol{k}\left(t'\right)\right]+\boldsymbol{F}_{IR}(t')\cdot\left\{\boldsymbol{\mathcal{A}}'_{g}\left[\boldsymbol{k}(t')\right]+\nabla_{\boldsymbol{k}}\phi_{d}\left[\boldsymbol{k}\left(t'\right)\right]\right\}\right]dt'
    \label{eq:action}
\end{equation}


where we used the property that any well-behaved function $f[\boldsymbol{k}(t)]$ satisfies \mbox{$\frac{d}{dt}f = -\boldsymbol{F}_{\mathrm{IR}}\cdot\nabla_{\boldsymbol{k}} f$} \cite{chacon2020circular}. We note that the last term in Eq. \ref{eq:action} describes the induced shift vector \cite{qian2022role}, accounting for the geometric shift between the electron and the hole wavefunctions, which is a physical gauge invariant and thus measurable quantity \cite{chacon2020circular}. The integral over this term is the induced interband Berry phase \cite{uzan-narovlansky_observation_2024}: $\gamma_B\equiv-\int_{t_{0}}^{t}\left[\boldsymbol{F}_{IR}(t')\cdot\left\{\boldsymbol{\mathcal{A}}'_{g}[\boldsymbol{k}(t')]+\nabla_{\boldsymbol{k}}\phi_{d}\left[\boldsymbol{k}\left(t'\right)\right]\right\}\right]dt'=\int_{\boldsymbol{k}_{0}}^{\boldsymbol{k}(t)}\boldsymbol{\mathcal{A}}'_{g}(\boldsymbol{k}')\cdot d\boldsymbol{k}'-\phi_{d}\left[\boldsymbol{k}\left(t_{0}\right)\right]+\phi_{d}\left[\boldsymbol{k}\left(t\right)\right]$. The Berry phase is in general a complex quantity \cite{berry1990geometric,whitney2005geometric,cohen2019geometric}, with the real and imaginary parts describing the geometric phase accumulated along the propagation and tunneling steps, respectively. Indeed, the tunneling event leads to a geometric amplitude attenuation or enhancement \cite{berry1990geometric}, while the propagation leads to the accumulation of a geometric phase. Overall, the Berry phase acts as a complex perturbation to the action, modifying both the amplitude and phase associated with each trajectory.


To resolve the complex Berry phase, we have to introduce an interferometric measurement, which captures the coherence of the electronic wavefunction on an attosecond time scale. This interferometer is realized by controlling two electron trajectories, driven by consecutive half-cycles of the IR field. Introducing a weak parallel-polarized second harmonic (SH) field perturbs these trajectories, allowing control over their complex properties. By precisely adjusting the delay between the IR and SH fields, we achieve accurate control over their relative amplitude and phase with attosecond resolution. As shown in Fig. 1b, the THz field displaces the electron wavepacket from the center of the lattice-site potential, effectively breaking inversion symmetry and enabling preferential tunneling to one side. The SH signal provides a sensitive probe of this symmetry breaking. By controlling the SH–IR delay, one can tune both the instantaneous shape of the tunneling barrier and the subsequent propagation path of the electron wavepacket, thereby controlling its trajectory length. This manipulation can either amplify or counteract the induced symmetry breaking, enhancing or suppressing its effect, while encoding the dynamical complex Berry phase. This interferometric scheme has been successfully applied to reveal a wide range of attosecond-scale phenomena, from recollision dynamics and field-induced tunneling in gas-phase systems \cite{pedatzur2015attosecond} to interband transitions \cite{vampa_linking_2015}, band structure dynamics \cite{uzan-narovlansky_observation_2022}, and the Berry phase in solids \cite{uzan-narovlansky_observation_2024,bai2024probing, faeyrman2024revealing}. In this study, the SH perturbation can act either constructively or destructively relative to the field-induced Berry phase, thereby isolating and revealing its complex contribution.


We experimentally induced a field-driven Berry phase by illuminating a 50-$\mu$m-thick MgO sample with a strong THz field and resolved the resulting Berry phase using two-color HHG interferometry. To break inversion symmetry and induce the Berry curvature, we applied a THz field with a peak amplitude of approximately 150 kV/cm, generated via optical rectification in a DAST organic crystal \cite{fulop2020laser}  . This field strength is typically sufficient to significantly break inversion symmetry in solids \cite{Vampa2018,Li2023}, as well as observe nonlinear THz-driven phenomena. The HHG process was driven by a strong, 45-fs-long IR pulse with a central wavelength of $\lambda$ = 1.33 $\mu$m, linearly polarized parallel to the THz field. The IR intensity on the sample was on the order of $10^{13}$ W/cm$^2$, generating harmonic emission extending up to 25 eV. The interferometric measurement is realized by a weak SH field with an intensity of approximately $1\%$ relative to the fundamental IR intensity, acting as a perturbation. The SH field was generated using a 100-$\mu$m-thick type-I phase-matching barium borate crystal (BaB$_2$O$_4$ - BBO). The relative delay between the IR and SH fields was precisely controlled using a pair of fused silica wedges mounted on a translation stage (see  details of the experimental setup in the SI).


To reveal symmetry breaking in the crystal, we first generate high-harmonic emission driven solely by the IR field and then compare the response in the absence and presence of the THz field. Figure 2a shows the harmonic spectrum measured in the absence of the THz field along the $\Gamma$-X orientation. As expected for a centrosymmetric material, only odd-order harmonics are observed. Next, we superimpose the THz field to manipulate the degree of symmetry breaking. This is achieved using a pair of wire-grid polarizers, with the THz intensity controlled by varying the relative angle between them (Fig. 2b). Adding the THz field leads to the emergence of even-order harmonics, as shown in Fig. 2c, indicating the breaking of inversion symmetry in the system. At first glance, the even harmonics do not appear clearly across the entire spectrum. However, a deeper examination, extracting the harmonic signal contrast as a function of the THz intensity modulations, reveals the even harmonics above the noise level across the entire energy range (Fig. 2d). Figures 2e (odd harmonics) and 2f (even harmonics) show the harmonic intensities as a function of the THz polarizer angle. As the THz field intensity increases, we observe a clear rise in even-harmonic intensity accompanied by a corresponding reduction in odd-harmonic intensity. This response confirms that the THz-induced symmetry breaking alters the interference between pathways contributing to odd and even harmonics.

Resolving the THz-field-induced Berry phase imposes two important challenges. First, we have to isolate the extremely weak symmetry breaking, accompanied by a high background signal. Second, we have to identify its complex properties, isolating both its real and imaginary components. We address these two challenges by adding a weak SH field, controlling the internal interferometer comprised of two consecutive electron trajectories. The SH field serves two important functions: it enhances the weak field-induced symmetry breaking signal and isolates its complex contribution. Specifically, adding the weak SH field perturbs both the asymmetry between half cycles during the tunneling process as well as the asymmetry in the electron trajectory during propagation (see Fig. 1b). These effects are incorporated into the semiclassical action as: $S=S_0+\gamma_B+\epsilon\sigma_{2\omega_0}$ with $\epsilon=\frac{F_{2\omega_0}}{F_{IR}}\ll1$, being the SH and IR fields amplitude ratio, $\omega_0$ the IR field central frequency, and $\sigma_{2\omega_0}$ is the SH perturbation. For the sake of clarity, we underline that each perturbation is a complex value and can be explicitly written in its real and imaginary parts as: $\sigma_{2\omega_0} = \sigma_{2\omega_0}^{r}+i\sigma_{2\omega_0}^{i}$, $\gamma_B=\gamma_B^r+i\gamma_B^i$. The imaginary parts reflect modifications to the electron’s dynamics during tunneling, whereas the real parts capture perturbations acquired during the propagation step \cite{pedatzur2015attosecond}. Due to symmetry, both perturbations invert signs between consecutive half cycles of the fundamental field, modulating the relative amplitude and phase of the emitted harmonics. We notice that the Berry phase is anti-symmetric with respect to the half-cycles (since $\boldsymbol{F}_{IR}$ is antisymmetric and due to TR symmetry $\boldsymbol{\mathcal{A}}_n$ is symmetric), while the dynamical phase is symmetric (due to TR symmetry $\varepsilon_g(-\boldsymbol{k})=\varepsilon_g^*(\boldsymbol{k})$). Since our interferometric scheme is sensitive only to anti-symmetric variations between half-cycles, it selectively detects geometric contributions while suppressing dynamical ones. This interference signal is then encoded in the intensity of the emitted even and odd harmonics \cite{pedatzur2015attosecond,uzan-narovlansky_observation_2022,vampa_merge_2017}. In this case, odd harmonics are proportional to $I_N^{odd}(\tau)\propto\bigl|\cos\left(\epsilon\sigma_{2\omega_0}\left(\tau\right)+\gamma_B\left(\tau_{THz}\right)\right)\bigr|^2$ and the even are proportional to $I_N^{even}(\tau)\propto\bigl|\sin\left(\epsilon\sigma_{2\omega_0}\left(\tau\right)+\gamma_B\left(\tau_{THz}\right)\right)\bigr|^2$
, where $\tau$ and $\tau_{THz}$ are the IR field delays with respect to the SH and the THz fields, respectively (see SI for full derivation). Expanding the harmonic intensity to second order in $\sigma_{2\omega_0}$ and to first order in $\gamma_B$ (since $\gamma_B\ll\sigma_{2\omega_0})$ leads us to the following expressions:

\begin{equation}
    I_{HHG}^{N}\left(\tau\right)\propto e^{-2Im\left\{ S_0\right\} }\cdot\begin{cases}
1-2\epsilon\sigma_{2\omega_0}^{r}\gamma_B^r+2\epsilon\sigma_{2\omega_0}^{i}\gamma_B^i-(\epsilon\sigma_{2\omega_0}^{r})^2+(\epsilon\sigma_{i,2\omega_0}^{i})^2 & N\ odd\\
2\epsilon\sigma_{2\omega_0}^{r}\gamma_B^r+2\epsilon\sigma_{2\omega_0}^{i}\gamma_B^i+(\epsilon\sigma_{2\omega_0}^{r})^2+(\epsilon\sigma_{i,2\omega_0}^{i})^2 & N\ even
\end{cases}
\label{eq:first_order_odd_even_intensity}
\end{equation}

where $\sigma^{r/i}_{2\omega_0}(\tau)$ oscillates in $2\omega_0$ and $\gamma_B^{r/i}$ oscillates slowly in $q\omega_{THz}$ with $\omega_{THz}$ being the THz frequency that interacts with the crystal. Equation \ref{eq:first_order_odd_even_intensity} reveals several insights. First, it shows that in the absence of the THz field, both the real and imaginary parts of the Berry phase, $\gamma_B^{r/i}$, vanish and the characteristic $4\omega_0$ oscillations observed in inversion-symmetric systems are recovered (i.e. the second order response $(\sigma^{r/i}_{2\omega_0})^2$ oscillates in $4\omega_0$)\cite{luu2018observing,vampa2015linking,uzan_attosecond_2020}. In this case, due to inversion symmetry, the emitted harmonic signal is equivalent for delays that are $T_{\omega_0}/4$ apart ($T_{\omega_0}$ being the fundamental field period), resulting in $4\omega_0$ oscillations of the harmonic signal as a function of the SH delay $\tau$. Secondly, we see that the first order contribution of $\sigma_{2\omega_0}^{r/i}$ is coupled to the THz induced Berry phase. 


Figure 3 presents experimentally resolved two-color interferometry of the complex Berry phase. Figure 3a presents the experimental setup of the two-color HHG combined with the THz-field induced dynamical symmetry breaking. This scheme allows for a 2D control over the two color-delay $\tau$ and the crystal orientation $\theta$. Figures 3b, 3e and 3h present the measured harmonics intensity oscillations recorded as a function of the SH delay for different crystal orientations with and without the THz perturbation for the three most intense harmonics (H16-H18). In the absence of the THz field, clear $4\omega_0$ oscillations are observed, as expected from an inversion symmetric material. Introducing the THz perturbation induces $2\omega_0$ oscillations alongside the intrinsic $4\omega_0$ component, providing a clear signature of symmetry breaking \cite{luu2018observing}. Figures 3d, 3g, and 3j present the Fourier spectra of the measured harmonic oscillations resolved along the $\Gamma-$X orientation. In the absence of the THz field (blue), the spectrum exhibits a pronounced peak at $4\omega_0$. When the THz field (orange) is applied, an additional peak appears at $2\omega_0$, revealing the inversion symmetry breaking. Figures 3c, 3f, and 3i present the corresponding numerical calculations. In the calculations, we include 12 bands of MgO, comprising 3 valence bands and 9 conduction bands, with a band gap of 7.8 eV. We adopt a Wannier-localized basis to eliminate gauge-related ambiguities in multiband SBE simulations, including random phase fluctuations, dipole discontinuities, and band degeneracies. The laser parameters are chosen to closely match the experimental conditions (see SI for a detailed description). A comparison between the numerical calculations and the experimentally measured results reveals excellent agreement.

Two-color HHG interferometry provides direct access to the complex Berry phase, enabling the isolation of its real and imaginary components, $\gamma_B^i$ and $\gamma_B^r$, across the harmonic spectrum. The reconstruction procedure is based on extracting the sum $I_{sum}=I_{HHG}^{N+1}+I_{HHG}^{N}$ and difference $I_{diff}=I_{HHG}^{N+1}-I_{HHG}^{N}$ between adjacent even and odd harmonics. The two observables separate the two complex components of the Berry phase, according to equation \ref{eq:first_order_odd_even_intensity}: $I_{sum}^{N}\propto1+4\gamma_B^i\sigma_{2\omega}^{i}+2(\sigma_{2\omega}^{i})^2$ and $I_{diff}^{N}\propto -1+4\gamma_B^r\sigma_{2\omega}^{r}+2(\sigma_{2\omega}^{r})^2$. We extract the $2\omega_0$ Fourier components of $I_{sum}$ and $I_{diff}$, normalized by their DC values, reconstructing the complex Berry phases across the HHG spectrum (see SI for the full derivation). The reconstructed complex phases encode the geometric properties of the electronic wavefunction, accumulated during the attosecond-scale dynamics. The imaginary Berry phase corresponds to the geometric phase accumulated under the tunneling barrier \cite{berry1990geometric, faeyrman2024revealing,zwanziger1991measuring}. It modulates the wavepacket amplitude in a geometric manner -- depending solely on the trajectory in the complex-valued parameter space and not on the system’s dynamics. Similarly, the real Berry phase is associated with the geometric phase accumulated by the electron-hole wavepacket while propagating in the conduction and valence bands, this time propagating in a real-valued parameter space \cite{uzan-narovlansky_observation_2024,bai2024probing}.



The above analysis reconstructs the complex Berry phase at a fixed delay of the THz field, determined by its instantaneous amplitude. The large separation between the IR and THz time scales enables access to Berry-phase dynamics within a single THz cycle. This separation justifies the adiabatic approximation, allowing the THz field to be treated as quasi-static at each delay point. Figure 4a illustrates the THz-field–induced Berry curvature that gives rise to a dynamical Berry phase. The oscillating THz field dynamically breaks inversion symmetry, generating a finite Berry curvature. As the field amplitude decreases and reverses sign, the Berry curvature correspondingly reduces and changes sign. Given the large separation between electronic and THz time scales, this modulation is effectively adiabatic from the electron’s perspective. The resulting transient Berry curvature imprints a geometric phase onto the electronic wavepacket, producing an induced transient Berry phase.

Experimentally, the relative delay between the IR and THz fields is varied during the IR–SH delay scan. Scanning the IR-SH delay, simultaneously shifts the THz–IR delay $\tau_{THz}$. Performing a careful Fourier analysis (see a detailed description in the SI) allows us to extract the temporal evolution of the real and imaginary components of the Berry phase. Figures 4b and 4c present the experimentally resolved temporal evolution of the field-induced Berry phases. Figure 4b describes the real components of the Berry phase, while Figure 4c presents the corresponding imaginary components for harmonics 14–17, associated with the first conduction band. The Berry phase exhibits temporal oscillations that directly reflect the degree of symmetry breaking induced by the THz field at each delay. These oscillations provide a dynamical fingerprint, revealing the evolution of the field-induced distortion of the crystal potential on sub-cycle timescales. Moreover, the real component of the Berry phase systematically exceeds its imaginary counterpart, pointing to the dominant role of the THz field during the intraband propagation step compared to tunneling. This interpretation is corroborated by SBE calculations, shown in Fig. 4d and 4e, which reproduce the experimental trend with good agreement.

\section{\textbf{Summary}}

We have demonstrated that an intense THz field can dynamically reshape the symmetry of a wide-bandgap insulator, giving rise to field-induced topology manifested through a field-driven Berry phase. To access this topology on attosecond timescales, we implemented a two-color HHG interferometric scheme. This approach disentangles the real and imaginary components of the Berry phase, resolving the field-induced topological phase accumulated during intraband propagation and the geometric amplitude modulation occurring at the tunneling step. The large separation between the relevant timescales enables us to follow the temporal evolution of the Berry phase within a single THz cycle. This measurement reveals the dynamical fingerprint of symmetry breaking, allowing the direct observation of field-driven dynamical geometrical phenomena in solids.

Our findings establish THz-driven solids, interrogated with attosecond-resolution HHG, as a powerful platform for tracking and controlling dynamical topological phenomena. In the future, such a scheme will enable access to hidden phases, ultrafast phonon-driven dynamics, and topological features otherwise inaccessible at equilibrium. This capability opens the door to a broad range of applications -- from engineering lightwave electronics based on transient topological states, to manipulating Berry curvature for valleytronics, and implementing coherent phase-control strategies relevant to quantum information processing. The ability to resolve and steer such phenomena on sub-cycle timescales makes THz-driven solids a unique laboratory for non-equilibrium quantum materials, where symmetry and topology become active degrees of freedom.

\newpage

\bibliography{main}
\bibliographystyle{unsrt}

\newpage
	\section*{Acknowledgments}
    We thank Alexander Poddubny for valuable theoretical insights regarding the impact of THz fields on inversion-symmetric systems. N.D. is the
incumbent of the Robin Chemers Neustein Professorial Chair.
N.D. acknowledges the Minerva Foundation, the Israeli Science Foundation and the European Research Council for the financial support. M.I. acknowledges funding of the DFG QUTIF grant IV152/6-2.

	\section*{Contributions}
    N.D. and R.P. conceived and designed the study. N.D., L.-Y.P, M.I. and R.P. supervised the project. R.P. and L.F. performed the experiments. L.F., J.N.Z., L.-Y.P and M.I. developed the theoretical framework. L.F. and R.P. analyzed the experimental data. J.N.Z., M.I. and L.-Y.P. developed the numerical code for the simulations. All authors discussed the results and contributed to writing the manuscript.
		
	\section*{Corresponding authors}
    Nirit Dudovich (nirit.dudovich@weizmann.ac.il) \\
    Riccardo Piccoli (riccardo.piccoli@unive.it)
	
    \section*{Data availability}
    The data and datasets that support the plots within this paper and other findings of this study are available from the corresponding author upon reasonable request.

        \section*{Code availability}
        The custom code used for the current study has been described in previous publications, and parts of it can be made available from the corresponding author on reasonable request.

        \section*{Supplementary information}
         Supplementary Information is available for this paper.

\newpage
\begin{figure}[hbt!]
	\centering
	  \includegraphics[width=0.95\textwidth]{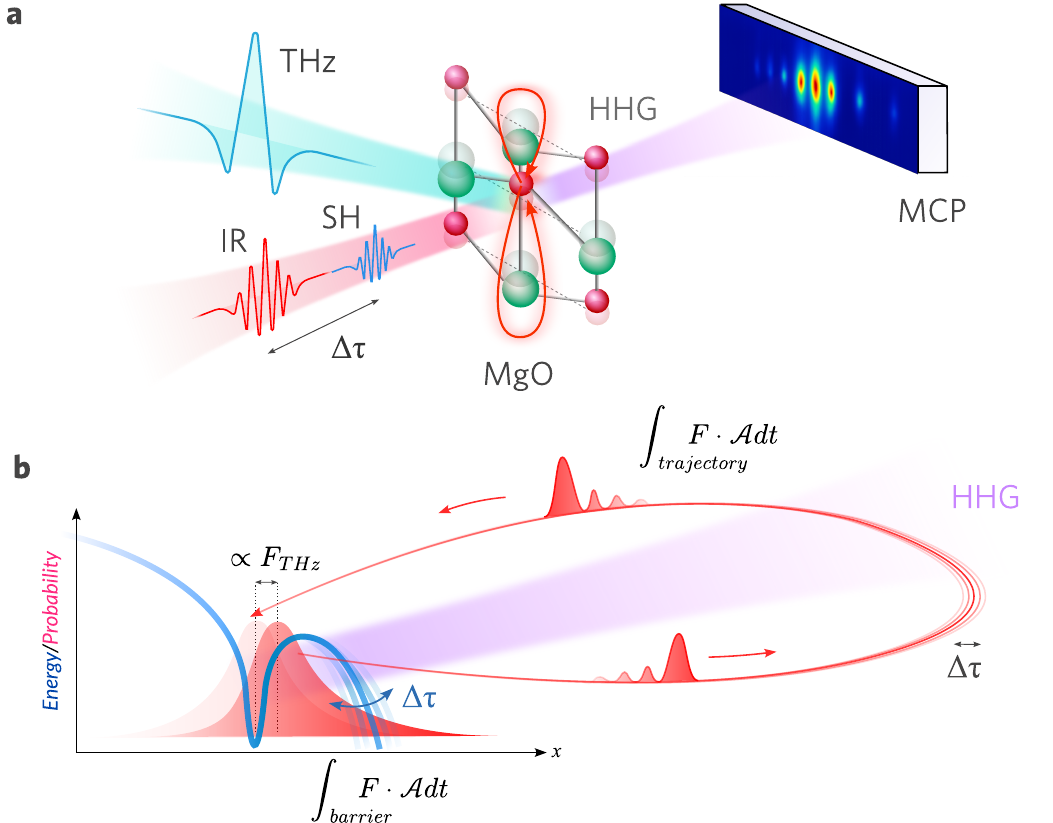}
\medskip
\caption{\textbf{Two-color HHG spectroscopy of THz-field induced Berry phase.} (\textbf{a}) A THz pulse (light blue) is focused onto the MgO crystal, inducing transient symmetry breaking. The transient Berry phase is probed via HHG generated by the fundamental IR laser field (red) and its SH (blue), and detected by a microchannel plate (MCP; blue screen on the right).
(\textbf{b}) Schematic illustration of the underlying mechanism. The initially symmetric lattice site potential (blue curve) is distorted by the fundamental IR field and further modulated by the SH field. In the presence of the THz field, the electron wavepacket (red) is displaced from the center of the lattice-site potential, effectively breaking inversion symmetry and enabling preferential tunneling to one side. This mechanism occurs throughout the crystal lattice. The SH delay controls both the modulation of the tunneling barrier ($\Delta\tau$, blue arrow) and the subsequent trajectory length during propagation ($\Delta\tau$, black arrow). As a result, different Berry phases are accumulated along the electron trajectories. Upon recombination, the emitted high-harmonic radiation encodes this accumulated Berry phase.}


	\label{fig1:fig1}
\end{figure}

\newpage
\begin{figure}[hbt!]
	\centering
	  \includegraphics[width=0.95\textwidth]{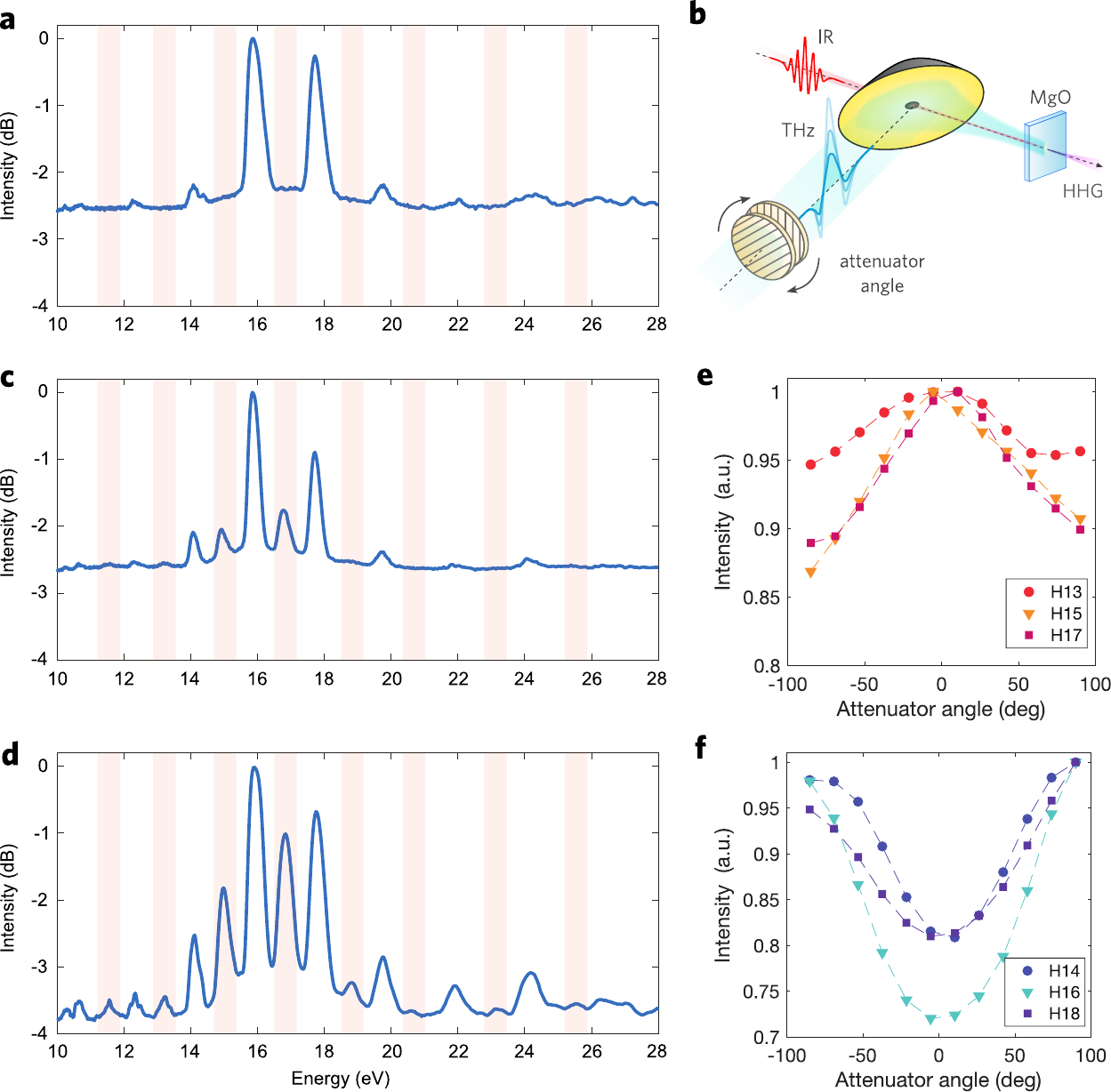}
\medskip
\caption{\textbf{Modulation of HHG spectra by the THz field.} (\textbf{a}) HHG spectrum obtained with the fundamental IR beam in the absence of the THz field. (\textbf{b}) Experimental setup: the THz radiation is focused onto the MgO sample together with the IR beam using a parabolic gold mirror. Two wire-grid polarizers are employed to control the THz field intensity as a function of their relative angle. (\textbf{c}) HHG spectrum showing the distinct emergence of even harmonics (e.g., near 15 and 17 eV) when the THz field is applied. (\textbf{d}) HHG spectrum as in (\textbf{c}), resolving the spectral components effectively modulated by the polarizer. With this analysis, all even harmonics become clearly visible across the entire energy range. Intensity of (\textbf{e}) odd and (\textbf{f}) even harmonics as a function of the relative angle between the polarizers. A pronounced out-of-phase behavior between even and odd harmonics is observed.}
	\label{fig2:fig2}
\end{figure}

\newpage
\begin{figure}[hbt!]
	\centering
	  \includegraphics[width=0.95\textwidth]{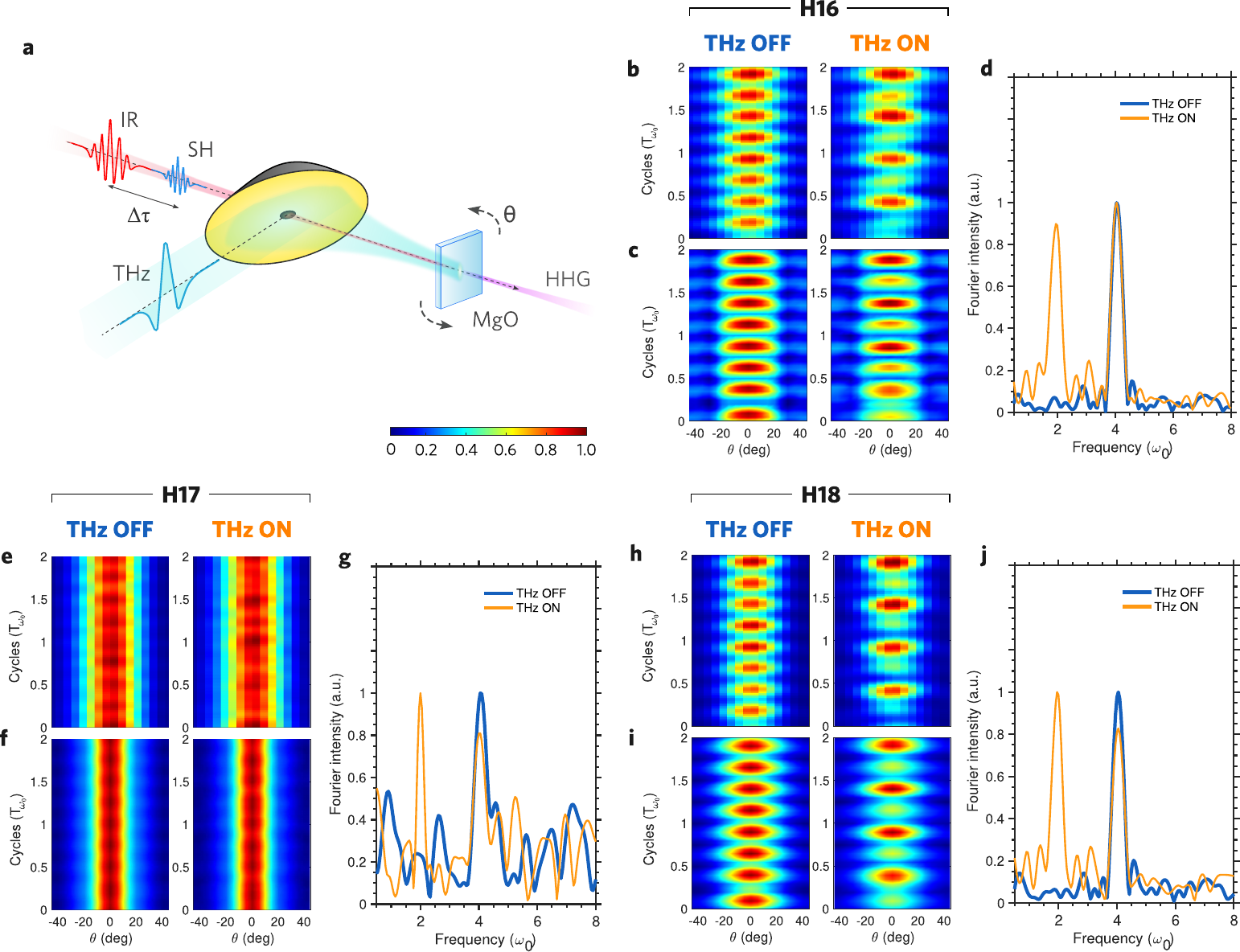}
      \medskip

      \caption{\textbf{THz modulated two-color HHG.} (\textbf{a}) Schematic description of the experimental setup. The THz field is focused onto the MgO sample together with the fundamental IR field and its SH, generating high harmonics. The 2D measurement resolves the HHG spectrum as a function of the IR-SH delay, $\tau$, and crystal orientation, $\theta$. (\textbf{b–j}) High-harmonic signal (a.u.) as a function of the two-color delay (vertical axis, in IR optical cycles) and crystal orientation (horizontal axis, degrees), for harmonics 16–18. Experimental measurement (\textbf{b,e,h}) and numerically simulation (\textbf{c,f,i}) of the 2D HHG, in the absence (left panels) and presence (right panels) of the THz field. The $\Gamma-X$ direction corresponds to 0 degrees. (\textbf{d,g,j}) Fourier transform over the two-color delay $\tau$ of the measured harmonic signal (a.u.) along the $\Gamma-X$ with and without the THz field.}

	\label{fig3:fig3}
\end{figure}

\newpage
\begin{figure}[hbt!]
	\centering
	  \includegraphics[width=0.9\textwidth]{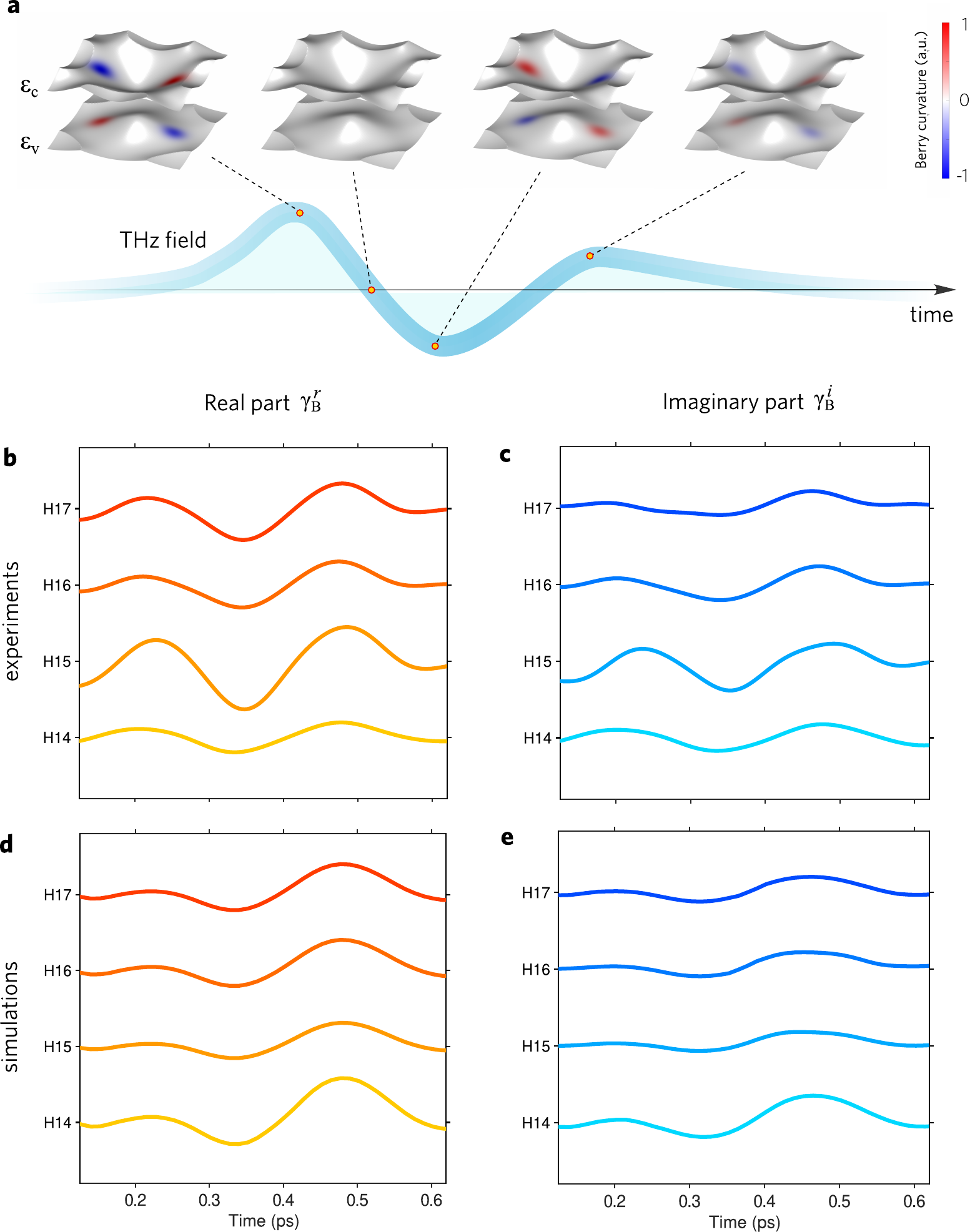}
\medskip
\caption{\textbf{Temporal evolution of the field-induced dynamical Berry phase.} (\textbf{a}) THz-field-induced Berry curvature. As the THz field breaks the crystal inversion symmetry, a non-vanishing Berry curvature emerges along the symmetry-breaking axis in the valence ($\varepsilon_v$) and conduction ($\varepsilon_c$) bands. The Berry curvature is superimposed on the band structure, with positive (negative) values shown in red (blue).
(\textbf{b–e}) Reconstructed temporal evolution of the real (\textbf{b,d}, yellow–red shades) and imaginary (\textbf{c,e}, cyan–blue shades) components of the field-induced Berry phase for harmonics 14–17 of the first conduction band, obtained from experimental (\textbf{b,c}) and simulated (\textbf{d,e}) data (in arbitrary units). The time axis is given in picoseconds. For clarity, the data for each harmonic are vertically offset.}

	\label{fig4:fig4}
\end{figure}

\end{document}